%Paper: hep-th/9204055
%From: TORRE@cc.usu.edu
%Date: Fri, 17 Apr 1992 09:17 MDT

%%                              JNL.TEX
%%
%%                This is JNL.TEX Version 0.3 as of 6/12/85.
%%
%%      This is a set of TeX 82 macros designed to produce scientific
%%      papers with a minimum of fuss and using as much of plain.tex as
%%      possible.  The user need only know what is in the TeXbook, and
%%      the macros under ``user definitions'' below.  Also, the user
%%      definitions are intended to be as simple as possible, so that
%%      the user may change them as desired.

%%
%%  Font definitions suitable for the IMAGEN (Written by Tony Kennedy)
%%

%  Define a whole menagerie of pseudo-12pt fonts

\font\twelverm=cmr10 scaled 1200    \font\twelvei=cmmi10 scaled 1200
\font\twelvesy=cmsy10 scaled 1200   \font\twelveex=cmex10 scaled 1200
\font\twelvebf=cmbx10 scaled 1200   \font\twelvesl=cmsl10 scaled 1200
\font\twelvett=cmtt10 scaled 1200   \font\twelveit=cmti10 scaled 1200

\skewchar\twelvei='177   \skewchar\twelvesy='60

%  Define \...point macros to change fonts and spacings consistently

\def\twelvepoint{\normalbaselineskip=12.4pt
  \abovedisplayskip 12.4pt plus 3pt minus 9pt
  \belowdisplayskip 12.4pt plus 3pt minus 9pt
  \abovedisplayshortskip 0pt plus 3pt
  \belowdisplayshortskip 7.2pt plus 3pt minus 4pt
  \smallskipamount=3.6pt plus1.2pt minus1.2pt
  \medskipamount=7.2pt plus2.4pt minus2.4pt
  \bigskipamount=14.4pt plus4.8pt minus4.8pt
  \def\rm{\fam0\twelverm}          \def\it{\fam\itfam\twelveit}%
  \def\sl{\fam\slfam\twelvesl}     \def\bf{\fam\bffam\twelvebf}%
  \def\mit{\fam 1}                 \def\cal{\fam 2}%
  \def\tt{\twelvett}
  \textfont0=\twelverm   \scriptfont0=\tenrm   \scriptscriptfont0=\sevenrm
  \textfont1=\twelvei    \scriptfont1=\teni    \scriptscriptfont1=\seveni
  \textfont2=\twelvesy   \scriptfont2=\tensy   \scriptscriptfont2=\sevensy
  \textfont3=\twelveex   \scriptfont3=\twelveex  \scriptscriptfont3=\twelveex
  \textfont\itfam=\twelveit
  \textfont\slfam=\twelvesl
  \textfont\bffam=\twelvebf \scriptfont\bffam=\tenbf
  \scriptscriptfont\bffam=\sevenbf
  \normalbaselines\rm}

%       tenpoint

%%
%%      Various internal macros
%%

\def\beginlinemode{\endmode
  \begingroup\parskip=0pt \obeylines\def\\{\par}\def\endmode{\par\endgroup}}
\def\beginparmode{\endmode
  \begingroup \def\endmode{\par\endgroup}}
\let\endmode=\par
{\obeylines\gdef\
{}}
\def\singlespace{\baselineskip=\normalbaselineskip}
\def\oneandathirdspace{\baselineskip=\normalbaselineskip
  \multiply\baselineskip by 4 \divide\baselineskip by 3}
\def\oneandahalfspace{\baselineskip=\normalbaselineskip
  \multiply\baselineskip by 3 \divide\baselineskip by 2}
\def\doublespace{\baselineskip=\normalbaselineskip \multiply\baselineskip by 2}

\newcount\firstpageno
\firstpageno=2
\footline={\ifnum\pageno<\firstpageno{\hfil}%
\else{\hfil\twelverm\folio\hfil}\fi}
\let\rawfootnote=\footnote              % We must set the footnote style
\def\footnote#1#2{{\rm\singlespace\parindent=0pt\rawfootnote{#1}{#2}}}
\def\raggedcenter{\leftskip=4em plus 12em \rightskip=\leftskip
  \parindent=0pt \parfillskip=0pt \spaceskip=.3333em \xspaceskip=.5em
  \pretolerance=9999 \tolerance=9999
  \hyphenpenalty=9999 \exhyphenpenalty=9999 }
\def\dateline{\rightline{\ifcase\month\or
  January\or February\or March\or April\or May\or June\or
  July\or August\or September\or October\or November\or December\fi
  \space\number\year}}
\def\received{\vskip 3pt plus 0.2fill
 \centerline{\sl (Received\space\ifcase\month\or
  January\or February\or March\or April\or May\or June\or
  July\or August\or September\or October\or November\or December\fi
  \qquad, \number\year)}}

%%
%%      Page layout, margins, font and spacing (feel free to change)
%%

\hsize=6.5truein
%\hoffset=1truein
\vsize=8.9truein
%\voffset=1truein
\parskip=\medskipamount
\twelvepoint            % selects twelvepoint fonts (cf. \tenpoint)
\oneandathirdspace           % selects double spacing for main part of paper
                        %      (cf. \singlespace, \oneandahalfspace)
\overfullrule=0pt       % delete the nasty little black boxes for overfull box

%%
%%      The user definitions for major parts of a paper (feel free to change)
%%

\def\preprintno#1{
 \rightline{\rm #1}}    % Preprint number at upper right of title page

\def\title                      %  Title on title page
  {\null\vskip 3pt plus 0.2fill
   \beginlinemode \doublespace \raggedcenter \bf}

\def\author                     %  Author(s) name(s)  on title page
  {\vskip 3pt plus 0.2fill \beginlinemode
   \singlespace \raggedcenter}

\def\affil                      % Affiliations (can intermix with \author)
  {\vskip 3pt plus 0.1fill \beginlinemode
   \oneandahalfspace \raggedcenter \sl}

\def\abstract                   % Begin abstract
  {\vskip 3pt plus 0.3fill \beginparmode
   \oneandathirdspace\narrower}

\def\endtitlepage               % End title page, begin body of paper
  {\endpage                     %       This subsumes \body
   \body}

\def\body                       % Begin text body;  can be used to end
  {\beginparmode}               % \title, \author, \affil, \abstract,
                                % \reference, or \figurecaption modes

\def\head#1{                    % Head;  NOTE enclose the text in {}
  \vskip 0.5truein     %  e.g., \head{I. Introduction}
  \noindent{\immediate\write16{#1}
   {\bf{#1}}\par}
   \nobreak\vskip 0.2truein\nobreak}

\def\subhead#1{                 % Subhead;  NOTE enclose the text in {}
  \vskip 0.25truein             % e.g., \subhead{A. History of the Problem}
  \noindent{{\it {#1}} \par}
   \nobreak\vskip 0.15truein\nobreak}

\def\refto#1{[#1]}           % For references in text as superscript

\def\references                 % Begin references -- basic format is Phys Rev
  {\subhead{References}         % I.e., volume, page, year (space after commas)
   \beginparmode
   \frenchspacing \parindent=0pt \leftskip=1truecm
   \oneandathirdspace\parskip=8pt plus 3pt \everypar{\hangindent=\parindent}}

\gdef\refis#1{\indent\hbox to 0pt{\hss#1.~}}    % Ref list numbers.

\gdef\journal#1, #2, #3, 1#4#5#6{               % Journal reference.  Comma set
    {\sl #1~}{\bf #2}, #3 (1#4#5#6)}           % off: name, vol, page, year

\def\refstylenp{                % Nucl Phys(or Phys Lett) ref style: V, Y, P
  \gdef\refto##1{ [##1]}                                % Reference in text []
  \gdef\refis##1{\indent\hbox to 0pt{\hss##1)~}}        % Ref list numbers)
  \gdef\journal##1, ##2, ##3, ##4 {                     % Journal reference
     {\sl ##1~}{\bf ##2~}(##3) ##4 }}

\def\refstyleprnp{              % Input like pr, output like np!!
  \gdef\refto##1{ [##1]}                                % Reference in text []
  \gdef\refis##1{\indent\hbox to 0pt{\hss##1)~}}        % Ref list numbers)
  \gdef\journal##1, ##2, ##3, 1##4##5##6{               % Journal reference
    {\sl ##1~}{\bf ##2~}(1##4##5##6) ##3}}

\def\pr{\journal Phys. Rev., }

\def\prd{\journal Phys. Rev. D, }

\def\jmp{\journal J. Math. Phys., }

\def\prpts{\journal Physics Reports, }

\def\cqg{\journal Class. Quantum Grav., }

\def\ann{\journal Ann. Phys., }

\def\endreferences{\body}

\def\figurecaptions             % Begin figure captions
  { \beginparmode
   \subhead{Figure Captions}
}

\def\endpage                    %  Eject a page
  {\vfill\eject}

\def\endpaper                   %  Ways to say goodbye
  {\endmode\vfill\supereject}

\def\endit
  {\endpaper\end}

%%
%%      Various little user definitions
%%

\def\ref#1{Ref. #1}                     %       for inline references
\def\Ref#1{Ref. #1}                     %       ditto

          % For citation of equation numbers
        %       ditto
                     %       ditto
                     %       ditto
                   %       ditto
                   %       ditto
\def\frac#1#2{{\textstyle{#1 \over #2}}}

\def\eg{{\it e.g.,\ }}

\def\ie{{\it i.e.,\ }}

\def\sla{\raise.15ex\hbox{$/$}\kern-.57em}
\def\leaderfill{\leaders\hbox to 1em{\hss.\hss}\hfill}
\def\twiddle{\lower.9ex\rlap{$\kern-.1em\scriptstyle\sim$}}
\def\bigtwiddle{\lower1.ex\rlap{$\sim$}}
\def\gtwid{\mathrel{\raise.3ex\hbox{$>$\kern-.75em\lower1ex\hbox{$\sim$}}}}
\def\ltwid{\mathrel{\raise.3ex\hbox{$<$\kern-.75em\lower1ex\hbox{$\sim$}}}}
\def\square{\kern1pt\vbox{\hrule height 1.2pt\hbox{\vrule width 1.2pt\hskip 3pt
   \vbox{\vskip 6pt}\hskip 3pt\vrule width 0.6pt}\hrule height 0.6pt}\kern1pt}

\def\prim{{\scriptscriptstyle{\prime}}}

\def\m@th{\mathsurround=0pt }
\def\leftrightarrowfill{$\m@th \mathord\leftarrow \mkern-6mu
 \cleaders\hbox{$\mkern-2mu \mathord- \mkern-2mu$}\hfill
 \mkern-6mu \mathord\rightarrow$}
\def\overleftrightarrow#1{\vbox{\ialign{##\crcr
     \leftrightarrowfill\crcr\noalign{\kern-1pt\nointerlineskip}
     $\hfil\displaystyle{#1}\hfil$\crcr}}}

%% *********** New stuff follows *******************

\font\titlefont=cmr10 scaled\magstep3

\def\martinstyletitle                      %  Title on title page
  {\null\vskip 3pt plus 0.2fill
   \beginlinemode \doublespace \raggedcenter \titlefont}

\font\twelvesc=cmcsc10 scaled 1200

\def\author                     %  Author(s) name(s)  on title page
  {\vskip 3pt plus 0.2fill \beginlinemode
   \singlespace \raggedcenter\twelvesc}

%%
%%      AmSTeX compatability definitions
%%
%%      To run a TeX file originally intended for AmSTeX, only small changes
%%      should be necessary (I hope).  Use the line \input jnl at the start.
%%      Remove the lines \input amstex, \documentstyle{itpjnl} at the
%%      beginning;  also remove all the page layout stuff (\parindent=1cm,
%%      \hsize=5.28125in etc.)  The page layout is now done automatically.
%%      Also OMIT the qualifier \magnification=1200 when you IMPRINT the
%%      .dvi file.  (\TagsOnRight is harmless, you can take it out or leave
%%      it in.)  I believe most AmSTeX will work with no change.  One problem
%%      is \footnote, which is a little different in that it now needs to
%%      have an explicit asterisk *  (or whatever) included, like this:
%%              \footnote*{Text winds up at bottom of page.}
%%      This is discussed on p. 116 of the TeXbook.  IGNORE the AmSTeX
%%      documentation (if you can call it that);  refer to the TeXbook.
%%
%%      Note that many commands in AmSTeX have their equivalents in the
%%      TeXbook, perhaps with different names and slightly differing
%%      usage. E.g., the old \align in AmSTeX is replaced by \eqalign
%%      (p. 190) and \aligntag is replaced by \eqalignno (p. 192).
%%      \align and \aligntag still work, but I recommend that you use
%%      \eqalign and \eqalignno in documents run under jnl.
%%
%%      See me if you have any problems  -- Doug.
%%

\def\heading                            % Heading
  {\vskip 0.5truein plus 0.1truein      % e.g., \heading I. NOTES \endheading
   \beginparmode \def\\{\par} \parskip=0pt \singlespace \raggedcenter}

\def\subheading                         % Subheading
  {\vskip 0.25truein plus 0.1truein     % e.g., \subheading{A. The Problem}
   \beginlinemode \singlespace \parskip=0pt \def\\{\par}\raggedcenter}

\def\tag#1$${\eqno(#1)$$}

\def\align#1$${\eqalign{#1}$$}

\def\aligntag#1$${\gdef\tag##1\\{&(##1)\cr}\eqalignno{#1\\}$$
  \gdef\tag##1$${\eqno(##1)$$}}

\def\endaligntag{}

\def\overset #1\to#2{{\mathop{#2}\limits^{#1}}}
\def\underset#1\to#2{{\let\next=#1\mathpalette\undersetpalette#2}}
\def\undersetpalette#1#2{\vtop{\baselineskip0pt
\ialign{$\mathsurround=0pt #1\hfil##\hfil$\crcr#2\crcr\next\crcr}}}

%%
%%      Various little user definitions
%%

\def\ref#1{Ref.~#1}                     %       for inline references
\def\Ref#1{Ref.~#1}                     %       ditto
\def\[#1]{[\cite{#1}]}
\def\cite#1{{#1}}
\def\(#1){(\call{#1})}
\def\call#1{{#1}}
\def\taghead#1{}
\def\frac#1#2{{#1 \over #2}}

\def\12{{1\over2}}
\def\eg{{\it e.g.,\ }}

\def\ie{{\it i.e.,\ }}

\def\sla{\raise.15ex\hbox{$/$}\kern-.57em}
\def\leaderfill{\leaders\hbox to 1em{\hss.\hss}\hfill}
\def\twiddle{\lower.9ex\rlap{$\kern-.1em\scriptstyle\sim$}}
\def\bigtwiddle{\lower1.ex\rlap{$\sim$}}
\def\gtwid{\mathrel{\raise.3ex\hbox{$>$\kern-.75em\lower1ex\hbox{$\sim$}}}}
\def\ltwid{\mathrel{\raise.3ex\hbox{$<$\kern-.75em\lower1ex\hbox{$\sim$}}}}
\def\square{\kern1pt\vbox{\hrule height 1.2pt\hbox{\vrule width 1.2pt\hskip 3pt
   \vbox{\vskip 6pt}\hskip 3pt\vrule width 0.6pt}\hrule height 0.6pt}\kern1pt}
\def\tdot#1{\mathord{\mathop{#1}\limits^{\kern2pt\ldots}}}

\def\pmb#1{\setbox0=\hbox{#1}%
  \kern-.025em\copy0\kern-\wd0
  \kern  .05em\copy0\kern-\wd0
  \kern-.025em\raise.0433em\box0 }

\catcode`@=11
\newcount\tagnumber\tagnumber=0

\immediate\newwrite\eqnfile
\newif\if@qnfile\@qnfilefalse
\def\write@qn#1{}
\def\writenew@qn#1{}
\def\w@rnwrite#1{\write@qn{#1}\message{#1}}
\def\@rrwrite#1{\write@qn{#1}\errmessage{#1}}

\def\taghead#1{\gdef\t@ghead{#1}\global\tagnumber=0}
\def\t@ghead{}

\expandafter\def\csname @qnnum-3\endcsname
  {{\t@ghead\advance\tagnumber by -3\relax\number\tagnumber}}
\expandafter\def\csname @qnnum-2\endcsname
  {{\t@ghead\advance\tagnumber by -2\relax\number\tagnumber}}
\expandafter\def\csname @qnnum-1\endcsname
  {{\t@ghead\advance\tagnumber by -1\relax\number\tagnumber}}
\expandafter\def\csname @qnnum0\endcsname
  {\t@ghead\number\tagnumber}
\expandafter\def\csname @qnnum+1\endcsname
  {{\t@ghead\advance\tagnumber by 1\relax\number\tagnumber}}
\expandafter\def\csname @qnnum+2\endcsname
  {{\t@ghead\advance\tagnumber by 2\relax\number\tagnumber}}
\expandafter\def\csname @qnnum+3\endcsname
  {{\t@ghead\advance\tagnumber by 3\relax\number\tagnumber}}

\def\equationfile{%
  \@qnfiletrue\immediate\openout\eqnfile=\jobname.eqn%
  \def\write@qn##1{\if@qnfile\immediate\write\eqnfile{##1}\fi}
  \def\writenew@qn##1{\if@qnfile\immediate\write\eqnfile
    {\noexpand\tag{##1} = (\t@ghead\number\tagnumber)}\fi}
}

\def\callall#1{\xdef#1##1{#1{\noexpand\call{##1}}}}
\def\call#1{\each@rg\callr@nge{#1}}

\def\each@rg#1#2{{\let\thecsname=#1\expandafter\first@rg#2,\end,}}
\def\first@rg#1,{\thecsname{#1}\apply@rg}
\def\apply@rg#1,{\ifx\end#1\let\next=\relax%
\else,\thecsname{#1}\let\next=\apply@rg\fi\next}

\def\callr@nge#1{\calldor@nge#1-\end-}
\def\callr@ngeat#1\end-{#1}
\def\calldor@nge#1-#2-{\ifx\end#2\@qneatspace#1 %
  \else\calll@@p{#1}{#2}\callr@ngeat\fi}
\def\calll@@p#1#2{\ifnum#1>#2{\@rrwrite{Equation range #1-#2\space is bad.}
\errhelp{If you call a series of equations by the notation M-N, then M and
N must be integers, and N must be greater than or equal to M.}}\else %
{\count0=#1\count1=#2\advance\count1 by1\relax\expandafter\@qncall\the\count0,%
  \loop\advance\count0 by1\relax%
    \ifnum\count0<\count1,\expandafter\@qncall\the\count0,%
  \repeat}\fi}

\def\@qneatspace#1#2 {\@qncall#1#2,}
\def\@qncall#1,{\ifunc@lled{#1}{\def\next{#1}\ifx\next\empty\else
  \w@rnwrite{Equation number \noexpand\(>>#1<<) has not been defined yet.}
  >>#1<<\fi}\else\csname @qnnum#1\endcsname\fi}

\let\eqnono=\eqno
\def\eqno(#1){\tag#1}
\def\tag#1$${\eqnono(\displayt@g#1 )$$}

\def\aligntag#1\endaligntag
  $${\gdef\tag##1\\{&(##1 )\cr}\eqalignno{#1\\}$$
  \gdef\tag##1$${\eqnono(\displayt@g##1 )$$}}

\def\eqalignno#1{\displ@y \tabskip\centering
  \halign to\displaywidth{\hfil$\displaystyle{##}$\tabskip\z@skip
    &$\displaystyle{{}##}$\hfil\tabskip\centering
    &\llap{$\displayt@gpar##$}\tabskip\z@skip\crcr
    #1\crcr}}

\def\displayt@gpar(#1){(\displayt@g#1 )}

\def\displayt@g#1 {\rm\ifunc@lled{#1}\global\advance\tagnumber by1
        {\def\next{#1}\ifx\next\empty\else\expandafter
        \xdef\csname @qnnum#1\endcsname{\t@ghead\number\tagnumber}\fi}%
  \writenew@qn{#1}\t@ghead\number\tagnumber\else
        {\edef\next{\t@ghead\number\tagnumber}%
        \expandafter\ifx\csname @qnnum#1\endcsname\next\else
        \w@rnwrite{Equation \noexpand\tag{#1} is a duplicate number.}\fi}%
  \csname @qnnum#1\endcsname\fi}

\def\ifunc@lled#1{\expandafter\ifx\csname @qnnum#1\endcsname\relax}

\let\@qnend=\end\gdef\end{\if@qnfile
\immediate\write16{Equation numbers written on []\jobname.EQN.}\fi\@qnend}

\catcode`@=12

\catcode`@=11
\newcount\r@fcount \r@fcount=0
\newcount\r@fcurr
\immediate\newwrite\reffile
\newif\ifr@ffile\r@ffilefalse
\def\w@rnwrite#1{\ifr@ffile\immediate\write\reffile{#1}\fi\message{#1}}

\def\writer@f#1>>{}
\def\referencefile{%			  Stuff to write .REF file
  \r@ffiletrue\immediate\openout\reffile=\jobname.ref%
  \def\writer@f##1>>{\ifr@ffile\immediate\write\reffile%
    {\noexpand\refis{##1} = \csname r@fnum##1\endcsname = %
     \expandafter\expandafter\expandafter\strip@t\expandafter%
     \meaning\csname r@ftext\csname r@fnum##1\endcsname\endcsname}\fi}%
  \def\strip@t##1>>{}}

\def\citeall#1{\xdef#1##1{#1{\noexpand\cite{##1}}}}
\def\cite#1{\each@rg\citer@nge{#1}}	% Variable No. of args, separated by

\def\each@rg#1#2{{\let\thecsname=#1\expandafter\first@rg#2,\end,}}
\def\first@rg#1,{\thecsname{#1}\apply@rg}	% each@ag is a general purpose
\def\apply@rg#1,{\ifx\end#1\let\next=\relax%	  variable no. of arg. macro.
\else,\thecsname{#1}\let\next=\apply@rg\fi\next}% args separated by commas

\def\citer@nge#1{\citedor@nge#1-\end-}	% Check for M-N range (M and N numbers)
\def\citer@ngeat#1\end-{#1}
\def\citedor@nge#1-#2-{\ifx\end#2\r@featspace#1 % Single argument
  \else\citel@@p{#1}{#2}\citer@ngeat\fi}	% M-N range of arguments
\def\citel@@p#1#2{\ifnum#1>#2{\errmessage{Reference range #1-#2\space is bad.}%
    \errhelp{If you cite a series of references by the notation M-N, then M and
    N must be integers, and N must be greater than or equal to M.}}\else%
 {\count0=#1\count1=#2\advance\count1 by1\relax\expandafter\r@fcite\the\count0,
  \loop\advance\count0 by1\relax%	  Loop from M to N
    \ifnum\count0<\count1,\expandafter\r@fcite\the\count0,%
  \repeat}\fi}

\def\r@featspace#1#2 {\r@fcite#1#2,}	% Eat spaces at beginning or end of arg
\def\r@fcite#1,{\ifuncit@d{#1}%		  Cite individual reference
    \newr@f{#1}%
    \expandafter\gdef\csname r@ftext\number\r@fcount\endcsname%
                     {\message{Reference #1 to be supplied.}%
                      \writer@f#1>>#1 to be supplied.\par}%
 \fi%
 \csname r@fnum#1\endcsname}
\def\ifuncit@d#1{\expandafter\ifx\csname r@fnum#1\endcsname\relax}%
\def\newr@f#1{\global\advance\r@fcount by1%
    \expandafter\xdef\csname r@fnum#1\endcsname{\number\r@fcount}}

\let\r@fis=\refis			% Save old \refis, redefine
\def\refis#1#2#3\par{\ifuncit@d{#1}%      Use two params #2 #3 to strip blank
   \newr@f{#1}%
   \w@rnwrite{Reference #1=\number\r@fcount\space is not cited up to now.}\fi%
  \expandafter\gdef\csname r@ftext\csname r@fnum#1\endcsname\endcsname%
  {\writer@f#1>>#2#3\par}}

\def\ignoreuncited{%   redefine \refis if ignoring uncited references
   \def\refis##1##2##3\par{\ifuncit@d{##1}%
    \else\expandafter\gdef\csname r@ftext\csname r@fnum##1\endcsname\endcsname%
     {\writer@f##1>>##2##3\par}\fi}}

\def\r@ferr{\endreferences\errmessage{I was expecting to see
\noexpand\endreferences before now;  I have inserted it here.}}
\let\r@ferences=\references
\def\references{\r@ferences\def\endmode{\r@ferr\par\endgroup}}

\let\endr@ferences=\endreferences
\def\endreferences{\r@fcurr=0%		  Save old \endreferences, redefine
  {\loop\ifnum\r@fcurr<\r@fcount%	  Loop over refnum and produce text
    \advance\r@fcurr by 1\relax\expandafter\r@fis\expandafter{\number\r@fcurr}%
    \csname r@ftext\number\r@fcurr\endcsname%
  \repeat}\gdef\r@ferr{}\endr@ferences}

% Save old \endpaper, redefine it to write parting message.

\let\r@fend=\endpaper\gdef\endpaper{\ifr@ffile
\immediate\write16{Cross References written on []\jobname.REF.}\fi\r@fend}

\catcode`@=12

\citeall\refto		% These macros will generate citations
\citeall\ref		%
\citeall\Ref		%

\ignoreuncited
\preprintno{FTG-112-USU}\dateline
\title Covariant Phase Space Formulation of Parametrized Field Theories
\author C. G. Torre
\affil Department of Physics
Utah State University
Logan, UT  84322-4415
USA
\abstract
Parametrized field theories, which are generally covariant versions of
ordinary field theories, are studied from the point of view of the covariant
phase space:  the space of solutions of the field equations equipped with a
canonical (pre)symplectic structure.  Motivated by issues arising in general
relativity, we focus on: phase space representations of the spacetime
diffeomorphism group, construction of observables, and the relationship
between the canonical and covariant phase spaces.
\endtitlepage
\oneandathirdspace
\def\sperp{{\scriptscriptstyle\perp}}
\def\DiffMtxt{{\it Diff(}${\cal M}${\it)\/} }
\def\DiffM{{\it Diff(}{\cal M}{\it)\/} }

\def\diffMtxt{{\it diff(}$\cal M${\it)\/} }
\def\Ma{{\cal M}^\alpha}
\def\Mm{{\cal M}^\mu}
\def\gab{g_{\alpha\beta}}
\def\gmn{g_{\mu\nu}}
\def\psia{\psi^A}
\def\phia{\varphi^A}
\taghead{1.}
\head{1. Introduction}
One of the central features of Newtonian mechanics is the presence of an
absolute time:  a preferred foliation of Galilean spacetime.  Despite the
presence of a universal notion of time, it is still possible to formulate
dynamics
in terms of an arbitrary time parameter.  This is the ``parametrized''
formulation of mechanics, which is obtained by adjoining the Newtonian
time to the configuration variables of the mechanical system
\refto{Lanczos1970}.  The
resulting
formalism is often elegant but, given the existence of a preferred time,
never really necessary.

The need for a field-theoretic formalism which includes arbitrary notions
of time (and
space) becomes apparent when one studies dynamical theories
consistent with Einstein's general theory of relativity.  Here there is {\it
no}
preferred standard of time (or space), and it is usually best to keep this fact
manifest by never selecting such a standard.  This can be done by
including the gravitational field, in the guise of the spacetime metric,  as a
dynamical variable and keeping the resulting
``general covariance'' --- more precisely:  the spacetime diffeomorphism
covariance--- of the theory manifest. However, it is not necessary to add
new physics (gravitational dynamics) in order to achieve a generally
covariant formulation of a field theory.  It has been known for a long time
\refto{ADM1959, Dirac1964, Kuchar1976} that any field theory on a
fixed background
spacetime can be made generally covariant by adjoining suitable spacetime
variables to the configuration space of the theory in much the same way as
one does in the parametrized formulation of mechanics.  This
diffeomorphism covariant formulation of field theory is likewise called
``parametrized field theory''.

Parametrized field theory allows one to study field theory without
prejudicing the choice of time (space), and for this reason alone it is a
useful
tool (see, \eg \refto{Halliwell1991}).  Because parametrized theories are
generally
covariant, they also
serve as an important paradigm for the dynamics
of gravitation \refto{Kuchar1972}.  Indeed, general relativity is often
viewed as an ``already
parametrized'' field theory; if this point of view could be explicitly
implemented then one can solve some very basic problems
\refto{Kuchar1992} which are
especially troublesome for the program of canonical quantization of the
gravitational field.

A relatively unexplored formulation of Hamiltonian gravity is based on the
``covariant phase space'' \refto{Witten1987, Wald1990, Ashtekar1991b,
Henneaux1991}.  The covariant phase space is defined as the space
of solutions to
the equations of motion and thus has the virtue of preserving manifest
covariance.  Because the space of solutions admits a (pre)symplectic
structure, one can still employ sophisticated Hamiltonian methods to
formulate the quantization problem.
Thus it is of interest to try and apply covariant phase space methods to
study the canonical quantum theory of
gravity.    Given the importance of parametrized
field theory both as a paradigm for general relativity and as an elegant
formulation of field theories, it is worth examining such theories from the
point of view of the covariant phase space.  In particular, how do the
stubborn
problems of time and observables \refto{Kuchar1992} appear in the
covariant phase space
formulation of parametrized field theory?  Can we use the parametrized
field theory paradigm to better understand the covariant phase space of
general relativity?  This latter question is made more pressing since it has
been shown recently that, strictly speaking, the canonical (as opposed to
covariant) phase space structure of
general relativity cannot be identified with that of any parametrized field
theory \refto{CGT1992a}.  As we shall see, the covariant and canonical
approaches
to the phase space of parametrized theories are quite different, and hence it
is plausible that the parametrized field theory paradigm will be more
suitable in the
context of the covariant phase space formulation.

In this paper we will present the covariant phase space
formulation of a general parametrized field theory. In particular we will
address the issue of the action of the diffeomorphism group on the phase
space, which is a delicate problem in the conventional Hamiltonian
formulation \refto{Isham1985}, as well as the related issue of how to
construct ``observables''
in this formalism. In canonical gravity the construction of observables has
so far proved intractable, so it is useful to see how the covariant phase
space
approach handles this question.  Most important perhaps, we will spell out
in detail the (somewhat complicated) relationship between the covariant and
canonical phase space approaches to parametrized field theory by
comparing the phase spaces, group actions, and observables in each
formulation.  Presumably, these results will at least hint at the
corresponding results in general relativity.

There are some disadvantages associated with trying to give an
analysis which includes a ``general field theory''.  In particular,
if one tries
to give too broad a coverage of possible field theories, then the description
becomes quite opaque if only because of the notational difficulties.  Thus,
for simplicity, we make some simplifying assumptions about the field
theories being studied that, while perhaps violated in some very exceptional
cases, are typically valid.  One important exception to the previous
statement is that
we will not attempt to include parametrized gauge theories in our analysis.
There are a couple of reasons for this.  First, the structure of a
parametrized gauge theory is rather different from that of a theory without
any gauge invariances.  This is because the parametrized formulation of
non-gauge theories leads to a phase space formulation
that is well-behaved with respect to the diffeomorphism group of the
spacetime manifold, while the parametrized gauge theory brings in the
larger group of bundle automorphisms.  It is an interesting problem to find
a globally valid formulation of parametrized gauge theory, but we shall not
do it here.  At any rate, if one wants to use parametrized field theory to
understand general relativity, then the relevant group is the
diffeomorphism
group and gauge theories can thus be played down in importance (however,
see \refto{Ashtekar1991a}).  One of the other main assumptions we will
make is
designed to simplify the task of relating the covariant and canonical phase
spaces.  Specifically, we will identify the space of Cauchy data for the field
theory with the canonical phase space for the theory. Given a spacelike
(Cauchy)
hypersurface, the Cauchy data will be
assumed to be the fields and their normal Lie derivatives on that
surface\footnote*{There are, of course, important cases which violate this
assumption, \eg the Dirac field, but such field theories present no new
features in the context of the present investigation.}.
In practice this identification, which
is tantamount to identifying the tangent and cotangent bundles over the
space of field configurations, is
done in terms of metrics, both on spacetime and on the manifold of fields,
but it will be too cumbersome to try and make explicit the identification.

The plan of the paper is as follows.  The next section deals with a brief
summary of the salient features of parametrized field theories;  this
includes the usual canonical formulation.  \S3 provides a quick tour of the
covariant phase space formalism and applies it to parametrized field
theories.   Next, in \S4 and \S5, we turn to the representations of
the diffeomorphism group on phase space and the extraction of the
observables; both of these issues are simply and neatly treated using the
covariant phase space. The final section, \S6, is in many ways the most
interesting; it spells out the relationship between the covariant and
canonical phase space formulations.
\taghead{2.}
\head{2.  Parametrized field theories}
We consider a collection of fields, $\psi^A$, propagating on a globally
hyperbolic spacetime
$({\cal M}, g_{\alpha\beta})$ according to the extrema of the action
functional
$$S[\psi^A] = \int_{\cal M}{{\cal L}(g; \psi^A, \partial\psi^A)}\tag21$$
For simplicity we assume that the fields are non-derivatively
coupled to the background geometry, and that the Lagrangian only depends
on the fields and their first derivatives.   The equations of motion are
$${\delta S\over\delta\psi^A}=0.\tag22$$
Note that the solution space of this equation generally cannot admit an
action of the
spacetime diffeomorphism group, \DiffMtxt, because the metric is fixed.

The parametrization process enlarges the configuration space by the space
of diffeomorphisms from $\cal M$ to itself.  When dealing with these new
dynamical variables, denoted $X$, it will be convenient to work
with two copies of $\cal M$, ${\cal M}^\alpha$ and ${\cal M}^\mu$, and then
view $X\in {\it Diff(}{\cal M}{\it)}$
as a map from ${\cal M}^\mu$ to ${\cal M}^\alpha$,
$$X:{\cal M}^\mu\to {\cal M}^\alpha.$$
Thus one can think of $X$ as a field on ${\cal M}^\mu$ taking
values in ${\cal M}^\alpha$.  Tensors on ${\cal M}^\mu$ will be
distinguished by Greek
indices from the end of the alphabet, likewise tensors on ${\cal M}^\alpha$
will
have Greek indices from the beginning of the alphabet.  As an important
example, the metric on ${\cal M}^\alpha$ is $g_{\alpha\beta}$;
given $X\in\DiffM$ this metric can be
pulled back to ${\cal M}^\mu$
$$\eqalign{g_{\mu\nu}&=\left(X^*g\right)_{\mu\nu}\cr
&=X^\alpha_\mu X^\beta_\nu g_{\alpha\beta}\circ X,}\tag23$$
where $X^\alpha_\mu$ is the differential of the map $X$.

Given $X\in\DiffM$, the Lagrangian density defined on $\Ma$ can be
pulled back to $\Mm$.  This gives an action which can be considered as a
functional of both $\varphi^A:=X^*\psi^A$,  and
$X$:
$$S[X,\varphi^A]=\int_{\Mm}{{\cal L}(X^*g; \varphi^A, \partial\phia),
}.\tag24$$
Because the original action integral is unchanged by a diffeomorphism
acting on both $\gab$ and $\psia$, the action $S[\varphi^A, X]$ is invariant
with respect to changes of the diffeomorphism $X$ provided one takes
into account the change induced in $\phia=X^*\psi^A$.  This leads to
the fact that if
$${\delta S[\varphi^A, X]\over\delta\phia}=0,\tag25$$
then the equations
$${\delta S[\varphi^A, X]\over\delta X^\alpha}=0\tag26$$
are automatically satisfied.  This can be verified directly.  The equations
\(26)
are equivalent to
$$\nabla_\mu T^{\mu\nu}=0,\tag28$$
where
$$T^{\mu\nu}=-2g^{-{1\over2}}{\delta S[\phia,X]\over\delta
\gmn}.\tag29$$
As is well known, \(28) follows from \(25).

The redundancy of the Euler-Lagrange equations associated with
$S[\varphi^A, X]$ is a consequence of the invariance of the action
functional (2.4) with
respect to the pull-back action of diffeomorphisms on its arguments.  If
$\phi\in\DiffM$ then
$$S[\phi^*\phia,X\circ\phi]=S[\phia, X].\tag29a$$
Thus the space of solutions to \(25) and \(26) will admit an action (in fact
more than
one) of \DiffMtxt.  This will be discussed in more detail in \S 4.

The canonical phase space formulation of parametrized field theory is
developed in \refto{Kuchar1976}; here we simply summarize the needed
results.  If
$X$
and $\phia$ are viewed as a collection of fields on $\Mm$, then to pass to
the Hamiltonian formulation we need a foliation $Y$ of $\Mm$:
$$Y:R\times\Sigma\to \Mm,$$
where $\Sigma$ is the 3-manifold representing space.  Tensors on $\Sigma$ are
represented
via Latin indices.  Derivatives along $R$ are denoted with a dot.  For each
$t\in R$, $Y$ becomes an embedding, $Y_{(t)}:\Sigma\to \Mm$.  We
demand that the embedded hypersurface is spacelike,
which means that the normal $n_\mu$ to the hypersurface, defined by
$$Y^{\phantom{(t)}\mu}_{(t)a} n_\mu = 0,\tag210$$
is timelike.  Here $Y^{\phantom{(t)}\mu}_{(t)a}$ is the differential of
the map $Y_{(t)}$.  We will normalize $n_\mu$ to unity:
$$g^{\mu\nu}n_\mu n_\nu=-1.\tag211$$
An equivalent way to express the requirement that the leaves of the
foliation are spacelike is to demand that the metric induced on $\Sigma$,
$$\gamma_{ab}:=Y^{\phantom{(t)}\mu}_{(t)a}Y^{\phantom{(t)}\nu}_{(t
)b}\gmn\circ Y_{(t)},\tag212$$
is positive definite for each $t$.  Note that, given a metric on $\Mm$
(induced from the fixed metric on $\Ma$), $n_\mu$ and $\gamma_{ab}$
are fixed functionals of $Y_{(t)}$.

The configuration space of the canonical formalism consists of pairs
$(q^A,Q)$, where $q^A:=Y_{(t)}^*\phia$ and $Q:=X\circ
Y_{(t)}$, which are just the fields pulled back to a slice.  Note that $Q$
represents an embedding of $\Sigma$ into $\Ma$:
$$Q:\Sigma\to\Ma,$$
with normal
$$n_\alpha=X_\alpha^\mu n_\mu\circ X^{-1}.\tag212a$$
In \(212a) we have defined $X^\mu_\alpha$ via
$$\eqalign{X^\mu_\alpha X_\mu^\beta &= \delta^\beta_\alpha,\cr
X^\mu_\alpha X_\nu^\alpha &= \delta^\mu_\nu,}\tag$$
\ie $X^\mu_\alpha$ is the inverse to the differential of $X\in\DiffM$,
viewed as a map of the tangent space at $p\in M$ to that at $X(p)$.
The embedding $Q$ is spacelike; let $Q^\alpha_a$ be the differential of
$Q$, then
$$\eqalign{(Q^*g)_{ab}&= Q^\alpha_a Q^\beta_b \gab\circ Q\cr
&=X^\alpha_\mu Y^\mu_a X^\beta_\nu Y^\nu_b \gab\circ (X\circ
Y_{(t)})\cr
&=\gamma_{ab}.}\tag213$$
Thus the configuration space can be viewed as that of the fields $q^A$
on $\Sigma$ along with the set of spacelike embeddings of $\Sigma$ into
$\Ma$.

The Hamiltonian form of the action is a functional of curves in the phase
space $\Upsilon$, the phase space consisting of pairs $(q^A, \Pi_A)$, $( Q,
P)$ where
$\Pi_A$ and $P$ are
conjugate to $q^A$ and $Q$ respectively; it takes the form
$$S[q, \Pi; Q, P,N]=\int_{R\times\Sigma}\left(\Pi_A \dot q^A +
P_\alpha\dot Q^\alpha - N^\alpha H_\alpha\right).\tag214$$
Here $\dot Q^\alpha$ is the derivative with respect to the parameter $t$ of
a 1-parameter family of
embeddings and is geometrically a vector field on $\Ma$; $P_\alpha$ is
therefore a 1-form density of weight one on $\Ma$.
$N^\alpha$ are Lagrange multipliers enforcing the first-class
constraints
$$H_\alpha:=P_\alpha+h_\alpha\approx0.\tag215$$
The constraints identify the momentum conjugate to the embedding with
$h_\alpha$, which represents the energy-momentum flux of the fields
$\psia$ through the hypersurface defined by $Q:\Sigma\to \Ma$;
$h_\alpha$ is a functional of $q^A, \Pi_A$ and $Q$.
The energy-momentum flux can be decomposed into its components
normal and
tangential to the hypersurface embedded in $\Ma$:
$$h_\alpha=-n_\alpha h + Q^a_\alpha h_a,\tag216$$
where $h$ is the energy density of $\psia$ as measured by an observer
instantaneously at rest in the hypersurface and $h_a$ is the corresponding
momentum density.  $Q^a_\alpha$ lifts 1-forms on $\Sigma$ to 1-forms on
$\Ma$ restricted to the embedded hypersurface and is defined as
$$Q^a_\alpha:=\gamma^{ab}\gab Q^\beta_b.\tag217$$
\taghead{3.}
\head{3. The covariant phase space}
The covariant phase space approach to dynamics exploits the point of view
that the phase space of Hamiltonian mechanics is a symplectic
manifold.  It can be shown that {\it any} (local) action for a
dynamical system contains within it the definition of a presymplectic
structure on its critical points \refto{Henneaux1991}.  If there are no
gauge transformations
in the theory, then the presymplectic structure is a genuine symplectic
structure and one can thus formulate Hamiltonian dynamics on the
space of solutions to the equations of motion.  In relativistic theories
this leads to a manifestly covariant phase space description.

The action functional $S$ can be viewed as a scalar function on the
space ${\cal A}$ of all field histories,  $S:{\cal A}\to R$.  From this point
of view, the
variation of a field is a tangent vector ${\cal V}$ to this space.  The
first variation of the action then can be viewed as the action on ${\cal
V}$ of the exterior derivative of $S$:
$$\delta S = dS({\cal V}).\tag31$$
Now restrict attention to the submanifold\footnote\dag{For
simplicity
we
ignore the possibility that the space of solutions has singularities.}
$\Gamma\subset{\cal A}$
of solutions to the equations of motion, then the first
variation of the action reduces to a surface term at the (asymptotic)
boundary of the
spacetime $\cal M$ (for an explicit expression
see \refto{Ashtekar1991b, Henneaux1991}):
$$i^*dS({\cal V})=\int_{\partial {\cal M}} j^a({\cal
V})d\Sigma_a.\tag32$$
Here $i:\Gamma\to {\cal A}$ is the natural embedding of the space of
solutions into the space of all fields.  The surface term defines the
(pre)symplectic potential $\Theta_\Sigma$, which is a 1-form on
$\Gamma$, via
$$\Theta_\Sigma({\cal V})=\int_\Sigma j^a({\cal V})d\Sigma_a,\tag33$$
where $\Sigma$ is a Cauchy surface in $\cal M$.  For simplicity we will
use the same notation ($\Sigma$) to denote an abstract 3-dimensional
manifold as well as for its image after an embedding.  Whenever it is
necessary to distinguish the two we will work explicitly with the
embedding.  In \(33) $\cal V$ is a
tangent
vector to $\Gamma$, {\it i.e.}, it is a solution of the linearized equations of
motion.  Note that, in general,
$\Theta_\Sigma$ depends on the choice of $\Sigma$.    Denote as
$\Omega$ the closed 2-form on $\Gamma$ obtained as the
exterior derivative of $\Theta_\Sigma$:
$$\Omega({\cal V, W})=d\Theta_\Sigma({\cal V, W}).\tag34$$
Because $d^2S=0$, it can be seen from \(32) that $\Omega$ is independent
of the
choice of $\Sigma$\footnote*{If $\Sigma$ is not compact this is true only
with suitable
boundary conditions at spatial infinity.}.
$\Omega$ is the (pre)symplectic structure.

If there are no gauge symmetries in the theory, then the Hessian of the
Lagrangian is non-degenerate and one can pass directly to the Hamiltonian
form of
the action.  From this form of the action it can be seen that the
(pre)symplectic
structure defined on the covariant phase space $\Gamma$ and the usual
symplectic structure on the momentum phase space are equivalent.  In
particular $\Omega$ is non-degenerate in this case and is
thus a true symplectic structure.

If the action functional admits gauge transformations then $\Omega$
necessarily has degenerate directions.  A detailed proof of this can be found
in \refto{Wald1990}; it is worth sketching a simple proof here.   First, we
shall define
a
gauge transformation ${\cal
G}:\Gamma\to\Gamma$ as any suitably differentiable map of $\Gamma$
onto itself that has arbitrary support on $\cal M$.  By ``support'' we mean
the
region of spacetime for which the transformation of field values in that
region is not the identity.   The requirement of arbitrary support is crucial;
it guarantees that gauge transformations are, roughly speaking,
parametrized by arbitrary functions on $\cal M$.  Now consider a
1-parameter
family of gauge transformations ${\cal G}_s$ beginning at the identity.  If
such families of
transformations do not exist then the symplectic structure need not be
degenerate.  Infinitesimal gauge transformations correspond to certain
``pure gauge'' tangent vectors ${\cal Z}$ to $\Gamma$,
$${\cal Z}:={d{\cal G}_s\over ds}\bigg|_{s=0},\tag35$$
which, thought of as fields on spacetime, have arbitrary support on $\cal M$.
We want to show that for each pure
gauge tangent vector ${\cal Z}$, $\Omega({\cal V, \cal Z})=0$ for all
choices of $\cal V$.  To do this consider $\Omega({\cal V, \cal Z})$ and
$\Omega({\cal V, Z^\prim})$, where the pure gauge
solutions to the linearized equations, $\cal Z$ and $\cal Z^\prim$,  are
chosen to be
identical in some (arbitrarily small) neighborhood of the hypersurface
$\Sigma$ in
$\cal M$ used to evaluate $\Omega({\cal V, \cal Z})$, but let $\cal
Z^\prim$ vanish
on some other hypersurface.  Such a pure gauge solution $\cal Z^\prim$
can always be
found because of the requirement that ${\cal G}_s$ have arbitrary support.
Because
$\Omega$ is defined by an integral involving the fields and their
derivatives on $\Sigma$, it is clear that $\Omega({\cal V,
Z})=\Omega({\cal V, Z^\prim})$.  On the other hand, because $\Omega$
is actually independent of the choice of the hypersurface and $\cal
Z^\prim$ vanishes on some hypersurface, we see that $\Omega({\cal V,
Z^\prim})=0$,  and hence $\Omega({\cal V, Z})=0$ for all choices of
$\cal V$.  We see that to every infinitesimal gauge transformation
corresponds a degenerate direction for $\Omega$.

As usual, one can show that in the degenerate case $\Omega$ is the
pull-back to $\Gamma$ of a non-degenerate 2-form, $\omega$, on the {\it
reduced phase space} $\hat\Gamma$, which is the space of orbits in
$\Gamma$ of the group of gauge transformations.  $\hat \Gamma$ is thus
a symplectic manifold (possibly with singularities); functions on $\hat\Gamma$
are the ``observables'' of the theory.  As shown in
\refto{Wald1990, Henneaux1991}, this
definition of the reduced phase space and observables is formally equivalent to
other
standard definitions, \eg that coming from the Hamiltonian formulation on
the usual canonical momentum phase space.

Application of the covariant phase space formalism to parametrized field
theory is
relatively straightforward.  The phase space $\Gamma$ is the space of
solutions to \(25) and \(26).  A point in $\Gamma$ is a pair $(\phia,X)$
satisfying
these equations. Tangent
vectors ${\cal V}$ to $\Gamma$ at $(\phia,X)$ are pairs
$(\delta\phia,\delta
X^\alpha)$, where $ \delta\phia$ is a solution to the field equations which
are linearized off $\phia$,
$$\int_{{\cal
M}^\prim}{\delta^2S[\varphi,X]\over\delta\phia\delta\varphi^{\prim B}}
\delta\varphi^{\prim B}=0,\tag36$$
and $\delta X^\alpha$ is a vector field on $\Ma$ generating a 1-parameter
family of diffeomorphisms.

The symplectic potential takes the form
$$\Theta_\Sigma({\cal V})=\int_\Sigma\left(\Pi_A\delta\phia - n_\mu
T_\alpha^\mu\delta X^\alpha\right),\tag37$$
where $n_\mu$ is the unit normal to the hypersurface $\Sigma$,
$$\Pi_A={\partial {\cal L}\over \partial(\partial_\mu\phia)}
n_\mu,\tag38$$
and
$$T_\alpha^{\phantom{\alpha}\mu}:=X_\alpha^\nu T_\nu^\mu.\tag39$$
Note that $\Pi_A$ is precisely the momentum conjugate to $q^A$ and,
evidently, $-n_\mu T^\mu_\alpha=-h_\alpha$ is the momentum conjugate
to
$Q$ in agreement with the standard canonical approach.

We can now take the exterior derivative (on $\Gamma$) of $\Theta$ to get
the (pre)symplectic form $\Omega$.  The explicit structure of $\Omega$
depends on the specific form of the Lagrangian, but the general expression
is of the form
$$\eqalign{\Omega({\cal V}, \hat {\cal V})=-\int_{\Sigma\times
\Sigma^\prim}\bigg[&2{\delta\Pi_A\over\delta\varphi^{\prim B}}\delta
\varphi^{[A}\delta\hat\varphi^{\prim B]}+2{\delta h_\alpha\over\delta
X^{\prim\beta}}\delta X^{[\alpha}\delta\hat
X^{\prim\beta]}+{\delta
h_\alpha\over\delta\varphi^{\prim B}}\left(\delta
X^\alpha\delta\hat\varphi^{\prim B}-\delta \hat
X^\alpha\delta\varphi^{\prim
B}\right)\cr
&+{\delta\Pi_A\over\delta
X^{\prim\alpha}}\left(\delta\varphi^A\delta\hat X^{\prim\alpha}-
\delta\hat\varphi^A\delta
X^{\prim\alpha}\right)\bigg],}\tag39a$$
where we use the primes to distinguish fields at different spatial points (on
the same hypersurface).
$\Omega$ is independent of the choice of $\Sigma$.

{}From the general argument presented above, we know that $\Omega$ has a
degenerate direction for each infinitesimal gauge transformation of the
theory.  Assuming the original (unparametrized) theory had no gauge
invariances, the degeneracy of $\Omega$ will stem from the action of
infinitesimal diffeomorphisms on $\Gamma$.  It is easily verified that
given $\phi\in\DiffM$ and a solution $(\phia, X)$ to the equations \(25),
\(26),
then $(\phi^*\phia, X\circ\phi)$ also satisfies these equations; this is simply
the statement that the field equations are ``covariant''.  Now let ${\cal
Z}=(L_v\phia, L_vX)$ be the pure gauge vector
field arising from the
induced action on $\Gamma$ of a 1-parameter family of
diffeomorphisms $\phi_s$ of $\Mm$  generated by the vector field
$v^\mu$\footnote*{$L_v$ denotes the Lie derivative and is defined as
$L_v\phia={d\over ds}\phi_s^*\phia\bigg|_{s=0}$, and $L_vX={d\over
ds}X\circ\phi_s\bigg|_{s=0}$.}.  Then
it follows that $\Omega({\cal V, Z})=0$ $\forall {\cal V}$.  The reduced
phase space $\hat\Gamma$ is the space of orbits in $\Gamma$ of
\DiffMtxt.  Actually, at this stage
one has to make a choice.  To obtain a reduced symplectic manifold, it is
sufficient to pass to the space of orbits of the subgroup ${\it
Diff}\!_{\scriptscriptstyle0}({\cal
M}) \subset\DiffM$ that is the connected component of the identity.
{}From the point of view of dynamics as symplectic geometry, it requires
additional physical input to identify points related by ``large
diffeomorphisms'' in $\DiffM$/${\it Diff}\!_{\scriptscriptstyle0}({\cal
M})$.  We will
try to proceed in such a way that our results are independent of the choice
made here.    At any rate, a globally valid gauge
which represents $\hat\Gamma$ is to simply fix $X$, \eg $X=identity$,
and we
recover the original unparametrized description of the field theory on
$\Ma$.
\taghead{4.}
\head{4.  Representations of the diffeomorphism group}
There are two natural symplectic actions of \DiffMtxt on $\Gamma$, one is
a right action the other is a left action.  The right action of $\phi\in\DiffM$
is defined via
$$\Phi_{right}\cdot(\phia, X)=(\phi^*\phia, X\circ\phi).\tag45$$
Because $\Omega$ as defined in \(39a) is independent of the choice of
Cauchy surface, it is straightforward to verify that $\Phi_{right}$
preserves the presymplectic structure.

Unlike the conventional Hamiltonian formulation of a generally covariant
theory, the phase space representation of the Lie algebra  \diffMtxt cannot
be via the Poisson algebra of functions $F$ on $\Gamma$ because, in light
of the degeneracy of the presymplectic form, the
definition of such functions is trivial:
$$dF=\Omega({\cal Z},\cdot )=0.\tag41$$
The representation of \diffMtxt on $\Gamma$ is via the 1-parameter
subgroups of \DiffMtxt which are realized by vector fields on $\Mm$.
These vector fields induce the pure gauge vector fields on $\Gamma$:
$${\cal Z}=(L_v\phia,L_vX),\tag41a$$
Note that because $\Omega$ is closed and has
degeneracy directions $\cal Z$ it follows that
$$L_{\cal Z}\Omega=0,\tag42$$
which is the infinitesimal version of the fact that $\Phi_{right}$ preserves
$\Omega$.
Given a 2-parameter family of symplectic diffeomorphisms generated by
the two vector fields on $\Gamma$:  ${\cal Z}=(L_v\phia,L_vX)$ and
${\cal Z}^\prime=(L_w\phia,L_wX)$, it is a straightforward computation
to show that the Lie bracket
$$[{\cal Z, Z}^\prime]={\cal Z}^{\prime\prime},\tag43$$
where
$${\cal Z}^{\prime\prime}=(L_{[w,v]}\phia, L_{[w,v]}X).\tag44$$
Thus the commutator algebra {\it vect(}$\cal M${\it)} of vector fields on
$\Mm$ is
anti-homomorphically mapped into the commutator algebra of ($\Omega$
preserving) vector fields, {\it vect($\Gamma$)} on $\Gamma$.  Using the
standard anti-homomorphism from  \diffMtxt into {\it vect(}$\cal
M${\it)}, we obtain a
homomorphism from \diffMtxt into {\it vect($\Gamma$)}.

It is also possible to define a left action of \DiffMtxt on $\Gamma$ by
letting
the diffeomorphisms act on $\Ma$ and then using $X$ to pull the results
back to $\Mm$.  Thus, given $\phi\in\DiffM$, we obtain new points in
$\Gamma$ via
$$(\phia, X)\rightarrow(\tilde\phi^*\phia, X\circ\tilde\phi)\tag46$$
where $\tilde\phi=X^{-1}\circ\phi\circ X$.  Note that the left action of
\DiffMtxt on $X$ amounts to a new choice of $X$ via $X\to\phi\circ X$,
and this leads to a redefinition of $\phia$ in terms of $\psi^A$:
$\phia=(\phi\circ X)^*\psi^A$.

The left action of \DiffMtxt on $\Gamma$ is an anti-homomorphism from
\DiffMtxt into the group of (pre)symplectic diffeomorphisms of
$\Gamma$.
This
can also be seen infinitesimally, \ie at the level of Lie algebras.  Fix a
point
$(\phia,X)\in\Gamma$.  A vector
field $v^\alpha$ on $\Ma$ generating a 1-parameter family of
diffeomorphisms of $\Ma$ defines a vector field on $\Mm$ via
$$v^\mu=X^\mu_\alpha v^\alpha\circ X=(X^*v)^\mu.\tag47$$  Even
though $v^\mu$
so-defined is a ``q-number'' (or in the language of \refto{Wald1990}
generates a field
dependent local symmetry), it still leads to degenerate directions for
$\Omega$ through \(42).  As before, if we let ${\cal
Z}=(L_{X^*v}\phia,L_{X^*v}X)$ and ${\cal
Z}^\prime=(L_{X^*w}\phia,L_{X^*w}X)$, then the commutator of these
two vector fields is given by
$$[{\cal Z, Z}^\prime]={\cal Z}^{\prime\prime},\tag48$$
where
$${\cal Z}^{\prime\prime}=(L_{X^*[v,w]}\phia,
L_{X^*[v,w]}X).\tag49$$

The right action of \DiffMtxt on $\Gamma$ views $\Mm$ as fundamental
and
$(\phia, X)$ as simply a collection of fields on $\Mm$.  It is this action of
the diffeomorphism group which is directly available on the covariant
phase space of general relativity \refto{Ashtekar1991b}.  The key feature
of the right
action
that makes it viable in general relativity is that it does not a require a
split
of the phase space into non-dynamical variables $X$ and dynamical
variables $\phia$.  The left action on the other hand stems from the action
of \DiffMtxt on $\Ma$, and it is only by identifying $\Ma$ as the image of
$\Mm$ under the map $X$ that this action can be constructed.  The
left
action is quite natural from the point of view of the parametrized field
theory because it realizes the diffeomorphisms directly
on $\Ma$, which is essentially the goal of the parametrization process. It
is unknown how to achieve such an action in general relativity.  This would
require a clean separation between gauge variables and dynamical
variables, which is of course a long-standing problem in gravitation.
\taghead{5.}
\head{5.  Observables}
Because the symplectic structure is degenerate, in order to obtain a
conventional phase space description one must pass to the reduced phase
space $\hat\Gamma$, which can be identified with the space of
diffeomorphism equivalence classes of the fields $(\phia, X)$ that satisfy
the field equations.  Functions on the reduced phase space are the
``observables'' of the theory.  The observables can be represented as
functions on $\Gamma$ which are invariant under the (left or right)
action of \DiffMtxt described in the last section.

An important class of
observables is obtained from any ``constants of motion'' that the the field
theory for $\psi^A$ may admit.  More generally, if there exists a p-form
$\beta$ built from the fields $\psi^A$ (and the metric $g_{\alpha\beta}$)
that is closed when $\psi^A$
satisfies its equations of motion, then the integral $Q_g[\psi]$ of $\beta$
over a closed
p-dimensional submanifold $\sigma$,
$$Q_g[\psi]=\int_\sigma \beta,\tag51$$
is independent of the choice of $\sigma$ (within its homology class).
The subscript $g$ indicates that $Q$ will in general depend on the
metric on $\Ma$.  Pulling $\beta$ back to $\Mm$ via $X$ yields a closed
p-form $\beta^\prim=X^*\beta$ on $\Mm$ and an observable
$ Q^\prim[X,\varphi]:=Q_{X^*g}[\varphi]$.   To
see this, consider a diffeomorphism $\phi$ of $\Mm$ and let $\sigma$ be some
p-dimensional
sub-manifold of $\Mm$.  For simplicity, let us assume that we restrict our
attention to orientation preserving diffeomorphisms.  Then for any integral we
have the identity
$$\int_{\sigma}\phi^*\beta^\prim=\int_{\phi(\sigma)}\beta^\prim,\tag52$$
where $\phi(\sigma)$ is the image of $\sigma$ under the diffeomorphism.
Because $\beta^\prim$ is closed, the right-hand side of \(52) is independent
of
the choice of closed p-dimensional submanifold within the homology class of
$\sigma$, which is preserved by \DiffMtxt, hence we can replace
${\phi(\sigma)}$ with $\sigma$ to conclude:
$$\int_{\sigma}\phi^*\beta^\prim=\int_{\sigma}\beta^\prim.\tag53$$  Thus
$Q^\prime[X\circ\phi, \phi^*\varphi]=Q^\prim[X, \varphi]$, and
$Q^\prim$ is an observable.

Unfortunately, there is no guarantee that there are any such closed forms
for a typical field theory, and even if they exist there will
usually be only a finite number of them.  What is usually desired is a {\it
complete} set of observables that can serve (at least locally) as a set of
coordinates on $\hat\Gamma$.  For a field theory such a set is necessarily
infinite-dimensional.

One complete set of observables that is always available if the
unparametrized field theory has no gauge symmetries are the fields
$\psi^A$ themselves.  Let us exhibit these observables as functions on the
covariant phase space $\Gamma$.  Given a point $(\phia,X)\in\Gamma$,
we can obtain a collection of fields ${\cal O}^A$ on $\Ma$ via
$${\cal O}^A:=X_*\phia,\tag54$$
where $X_*$ denotes the push-forward of tensors on $\Mm$ to tensors on
$\Ma$ by $X$.  By the way we constructed the parametrized formalism in
\S2, it is clear that the fields ${\cal O}^A$, defined by \(54), satisfy the
equations
of motion \(22) and hence are identifiable with the fields $\psi^A$ of the
unparametrized theory.  Are the fields ${\cal O}^A$, viewed as functions
on
$\Gamma$, observables?  To see that they are we examine the
right action of \DiffMtxt on $\Gamma$ and verify that ${\cal O}^A$ is
invariant
under this action.  The right action of $\phi\in\DiffM$ on ${\cal O}^A$ is
$$\eqalign{\Phi_{right}\cdot{\cal O}^A&=(X\circ\phi)_*(\phi^*\phia)\cr
&=(X^{-1})^*(\phi^{-1})^*(\phi^*\phia)\cr
&=(X^{-1})^*\phia\cr
&=X_*\phia\cr
&={\cal O}^A}\tag55$$
It follows that ${\cal O^A}$ are also left invariant by the left action of
\DiffMtxt on $\Gamma$.
\taghead{6.}
\head{6. Relation to the canonical theory}
Let us now compare the canonical and covariant viewpoints on
the phase space, the gauge group, and the observables.
\subhead{\it Phase Space}
The covariant phase space $\Gamma$ is built from spacetime fields and
spacetime diffeomorphisms satisfying \(25), \(26).  The canonical phase
space
$\Upsilon$ is built from spatial fields  which are Cauchy data for \(25),
along with  spacelike embeddings and their conjugate momenta.  How can
$\Gamma$ and $\Upsilon$ be related?  Let us begin by
answering the question at the level of the unparametrized field theory
describing the fields $\psi^A$ on $\Ma$.   Assuming the Cauchy problem is
well-posed, there is a bijection between the space of solutions to \(22) and
the set of Cauchy data for \(22).  In fact, there are an infinite number of
ways
to construct a bijection from the space of Cauchy data onto the space of
solutions.  This can be seen as follows.  Introduce an
arbitrary---but fixed---spacelike hypersurface $\Sigma$ in $\Ma$.  Because
the Cauchy problem
is well-posed, each set of Cauchy data on $\Sigma$ leads to a unique
solution of \(22).  Conversely, each solution to \(22) induces a (unique) set
of
Cauchy data on $\Sigma$.  For each choice of $\Sigma$ such a
correspondence can be made;  each map between Cauchy data and
spacetime solutions is bijective provided the function spaces for the solution
space and Cauchy data are appropriately chosen.  The symplectic structures
on the covariant phase space and on the space of Cauchy data are mapped
into each other by the induced action of the bijection.  More succinctly, the
covariant phase space and the canonical phase space are symplectically
diffeomorphic.

Now return to the parametrized theory.  A point in $\Gamma$ is
determined by (i) picking a diffeomorphism, (ii) pulling back the
prescribed
metric on $\Ma$, (iii) solving the Euler-Lagrange equations \(25), which
are
defined in terms of the pulled back metric.  An allowed point in the
canonical phase space lies in the constraint surface $\bar\Upsilon$ defined
via \(215); a point in $\bar\Upsilon$ is obtained by simply picking a
spacelike
embedding
$Q:\Sigma\to\Ma$ and a set of Cauchy data on $\Sigma$ (the embedding
momenta are determined by the constraints \(215)).  Corresponding to a
given point in
$\Gamma$ there are an infinity of points in $\bar\Upsilon$ because for
every
spacelike embedding there is a set of Cauchy data which generates the given
solution.  In the formalism based on $\Upsilon$ it is precisely the
canonical transformations generated by the constraint functions in \(215)
which
map points in $\bar\Upsilon$ to other points in $\bar\Upsilon$
corresponding to the same
spacetime solution.  This redundancy in $\Upsilon$ is somehow to be
matched by
the redundancy in $\Gamma$, which treats diffeomorphically related
solutions as distinct.

The relation between $\Gamma$ and $\bar\Upsilon$ is
again made by introducing an embedding $Y:\Sigma\to\Mm$; for now we
will not assume that the embedded hypersurface is spacelike.   For each
diffeomorphism $X:\Mm\to\Ma$ there is an embedding
$X\circ Y$ of $\Sigma$ into $\Ma$.  In addition, the solutions to \(25) (and
their
derivatives) can be pulled back to $\Sigma$ using $Y$.  Thus each
point in $\Gamma$ defines a point in the
product of the space of Cauchy data for \(25) (or \(22)) and the space of
embeddings of
$\Sigma$ into $\Ma$.  Let us denote this product space as
$\Upsilon^\prim$ and the image of $\Gamma$ in $\Upsilon^\prim$ as
$\Lambda_Y$.  Note that the map from $\Gamma$
to $\Upsilon^\prim$ need not be surjective and certainly cannot be injective
because two different diffeomorphisms $X_1\neq
X_2$ can have the same action on a given hypersurface: $X_1\circ
Y=X_2\circ Y$, and two distinct solutions to the field equations \(25) can
induce
the same data on a slice provided the slice is not a Cauchy surface.

Because the space of {\it spacelike} embeddings is
an open submanifold of the space of embeddings, it follows that
$\bar\Upsilon$ is an open submanifold of $\Lambda_Y$.  The fact that the
constraint surface arising in the canonical approach can be identified as a
proper subset of the (image in $\Upsilon^\prim$ of the) covariant phase
space has important repercussions for the
action of the spacetime diffeomorphism group on the canonical phase
space.

Let us denote the inverse image of $\bar\Upsilon$ as
$\bar\Gamma$, and $\pi:\bar\Gamma\to\bar\Upsilon$ the surjection which
assigns to a point $(\phia, X)\in\bar\Gamma$ the point
$(q^A,p^A,Q)\in\bar\Upsilon$, where
$$\eqalign{q^A&= Y^*\phia,\cr
p^A&=Y^*L_n\phia,\cr
Q&=X\circ Y}.\tag61$$
The map $\pi$ is not injective.  To see why, let us think of
$\pi$ as taking a solution $\phia$ and a diffeomorphism $X$ and
constructing a spacelike embedding $Q:\Sigma\to\Ma$ and the Cauchy data
on this hypersurface for the solution $\psi^A=X_*\phia$.  This
interpretation is possible because (i) $X\circ Y:\Sigma\to\Ma$ is by
assumption a spacelike embedding and (ii) $Q^*X_*=Y^*$.  Now,
if two points in $\bar\Gamma$ are mapped to the same point in
$\bar\Upsilon$
then, because the Cauchy problem is well-posed, the two points in
$\bar\Gamma$
necessarily correspond to the same solution $\psi^A$.  Because of the way
the parametrized field theory is constructed from the field theory on
$\Ma$, or, equivalently, from our construction of observables in \S5, it is a
simple exercise to see that this can happen if and only if the two points in
$\bar\Gamma$ are
related by the (right) action of \DiffMtxt on $\Gamma$.  Thus $\pi$
fails to be injective
whenever (i) one has two diffeomorphisms $(X_1, X_2)$ which have the
same action on
the fiducial embedding $Y$, (ii) the two diffeomorphisms and two
corresponding solutions to \(25), $(\phia_1, \phia_2)$, are related by the
right action of (yet another)
diffeomorphism $\rho:\Mm\to\Mm$.  Notice that (i) and (ii) imply $\rho$
must necessarily fix the embedding $Y$:
$$\rho\circ Y=Y.\tag$$

Having spelled out the relationship between $\bar\Gamma$ and
$\bar\Upsilon$, let us relate the respective presymplectic structures.The
presymplectic potential on the constraint surface \(215)
$\bar\Upsilon\subset\Upsilon$ can be written as
$$\theta(\delta q^A,\delta\Pi_A,\delta Q)=\int_\Sigma\left(\Pi_A\delta q^A
- h_\alpha\delta Q^\alpha\right).\tag62$$
The map $\pi$ pushes forward a vector ${\cal V}
=  (\delta\phia, \delta X)$ tangent to $\bar\Gamma$ at $(\phia, X)$ to a
vector $\pi_*{\cal V}=(Y^*\delta\phia,
Y^*L_n\delta\phia, \delta X\circ Y)$ tangent to $\bar\Upsilon$ at
$(Y^*\phia, X\circ Y)$.
It follows from \(37) that on $\bar\Gamma$ we have $\Theta_\Sigma({\cal
V})=\theta(\pi_*{\cal V})$, and hence
$\Theta_\Sigma=\pi^*\theta$.   This means that, on $\bar\Gamma$,
$\Omega=d\Theta_\Sigma$ is the pull back by $\pi$ of the presymplectic
structure
$d\theta$ on the constraint surface in $\Upsilon$.  Note that while the
identification of $\bar\Gamma$ with $\bar\Upsilon$ is dependent on the
choice of $Y:\Sigma\to\Mm$, the presymplectic structure itself is
independent of the choice of $Y$.
\subhead{\it Gauge transformations}
We have exhibited both a left and a right action of \DiffMtxt on the
covariant
phase space $\Gamma$.  Because the map from $\Gamma$ to
$\Upsilon^\prim$ is neither one to one nor onto there is no reason to
expect that we can push forward to $\Upsilon^\prim$ the Hamiltonian
vector fields which generate the group action, and it is easy to check that in
fact we cannot carry the group action from $\Gamma$ to
$\Upsilon^\prim$.  However, if we restrict attention to
$\pi:\bar\Gamma\to\bar\Upsilon$ the situation improves.  It is still not
possible to push forward the vector fields generating the right action, but it
{\it is} possible to push forward the Hamiltonian vector fields generating
the left
action.  To see this, we must check that the failure of $\pi$ to be injective
does not destroy the induced group action on $\bar\Upsilon$.  Consider
two points $(\phia_1,X_1)$ and
$(\phia_2,X_2)$ in $\bar\Gamma$ which map to the same point $(Q, q^A,
p^A)$ in $\bar\Upsilon$.  Now consider an infinitesimal
diffeomorphism\footnote*{We use an infinitesimal diffeomorphism so as
to preserve the
spacelike character of the embeddings; the infinitesimal action of one
parameter subgroups is sufficient for studying the Hamiltonian vector
fields.} $\phi$ whose left action on $\bar\Gamma$ gives two new points
$((X_1^{-1}\circ\phi\circ X_1)^*\phia_1,\phi\circ X_1)$ and $((X_2^{-
1}\circ\phi\circ X_2)^*\phia_2,\phi\circ X_2)$.  The infinitesimal group
action carries over consistently to $\bar\Upsilon$ because, as mentioned
above,
there must exist $\rho\in\DiffM$ that fixes $Y$ and such that
$X_2=X_1\circ\rho$, $\phia_2=\rho^*\phia_1$.  In detail
$$\eqalign{Q&\to Q_1^\prim=\phi\circ
X_1\circ Y\cr
 q^A&\to q^{\prim A}_1=Y^*(X_1^{-1}\circ\phi\circ X_1)^*\phia_1\cr
p^A&\to p_1^{\prim A}=Y^*(X_1^{-
1}\circ\phi\circ X_1)^*L_n\phia_1}\tag$$
is consistent with
$$\eqalign{Q&\to Q_2^\prim=\phi\circ
X_2\circ Y\cr
 q^A&\to q^{\prim A}_2=Y^*(X_2^{-1}\circ\phi\circ X_2)^*\phia_2\cr
p^A&\to p_2^{\prim A}=Y^*(X_2^{-
1}\circ\phi\circ X_2)^*L_n\phia_2}\tag$$
because (by assumption)
$$Q_2^\prim=\phi\circ X_2\circ Y=\phi\circ X_1\circ Y=Q_1^\prim\tag$$
and
$$\eqalign{q_2^{\prim A}=Y^*(X_2^{-1}\circ\phi\circ X_2)^*\phia_2
&=((X_1\circ\rho)^{-1}\circ\phi\circ X_1\circ Y)^*\rho^*\phia_1\cr
&=Y^*(X_1^{-1}\circ\phi\circ X_1)^*\phia_1=q_1^{\prim A},}\tag$$
$$\eqalign{p_2^{\prim A}=Y^*(X_2^{-1}\circ\phi\circ X_2)^*L_n\phia_2
&=((X_1\circ\rho)^{-1}\circ\phi\circ X_1\circ Y)^*\rho^*L_n\phia_1\cr
&=Y^*(X_1^{-1}\circ\phi\circ X_1)^*L_n\phia_1=p_1^{\prim
A}.}\tag62a$$
In \(62a) we used the fact that $\rho$ leaves invariant the hypersurface
embedded by $Y$ so
that $(\rho^* n)^\mu=n^\mu$.

We see then that the infinitesimal left action of \DiffMtxt on $\Gamma$ can
be carried over to the canonical phase space formalism, \ie the Lie algebra
\diffMtxt is realized on $\bar\Upsilon$, which was shown completely
within
the canonical approach by Isham and Kucha\v r \refto{Isham1985}.  But
the group action fails to carry over for two reasons.  First, the failure of
$\pi$ to be 1-1 prevents the
Hamiltonian vector fields generating the action of \DiffMtxt on $\Gamma$
from being pushed forward to $\Upsilon^\prim$.  Second, while the vector
fields on $\bar\Gamma$ can be pushed forward to $\bar\Upsilon$, these
vector fields cannot be complete because (as emphasized in
\refto{Isham1985}) \DiffMtxt
always maps some spacelike hypersurface into one that is not spacelike.

Often one does not exhibit the constraint functions in the form \(215) as is
natural when studying the representation of \diffMtxt, but rather in the
projected form
$$\eqalign{H_\sperp &:= n^\alpha H_\alpha\approx 0\cr
H_a &:= Q^\alpha_a H_\alpha\approx0.}\tag64$$
When smeared with a scalar function $N^\sperp$ and a vector $\bf N$ on
$\Sigma$ , these
constraint functions respectively generate canonical transformations
corresponding to
normal and tangential deformations of the embedding of $\Sigma$ into
$\Ma$.  This is also the meaning of the constraints occurring in general
relativity although it is not known how to cast them into the
parametrized form.  In contrast with the deformation of a hypersurface
along some arbitrary vector field, a {\it normal} deformation involves the
metric
and this leads to the well-known complication that the Poisson algebra of
the projected constraint functions cannot represent a Lie algebra. Thus if
we define
$$\eqalign{H(N^\sperp)&:=\int_\Sigma N^\sperp H_\sperp\cr
H({\bf N})&:=\int_\Sigma N^a H_a,}\tag$$
then we have the Poisson brackets
$$\eqalign{\left[H(N^\sperp),H(M^\sperp)\right]&=H\left({\bf J}\right)\cr
\left[H({\bf N}),H({\bf M})\right]&=H\left(L_{\bf N}{\bf M}\right)\cr
\left[H(N^\sperp),H({\bf M})\right]&=H\left(-L_{\bf
M}N^\sperp\right),}\tag65$$
 where
$$J^a=\gamma^{ab}\left(N^\sperp\partial_bM^\sperp-
M^\sperp\partial_bN^\sperp
\right),\tag66$$
and hence
the finite (as opposed to infinitesimal) canonical transformations generated
by the projected
constraint functions cannot realize a Lie group.  The ``open algebra'' \(65)
can be summarized in terms of functions $H(N^\sperp,{\bf N})$ on
$\Upsilon$, where
$$H(N^\sperp,{\bf N}):=H(N^\sperp) + H({\bf N}),\tag66a$$
satisfies
$$[H(N^\sperp,{\bf N}), H(M^\sperp,{\bf M})]=H(L_{\bf N}M^\sperp-
L_{\bf M}N^\sperp,L_{\bf N}{\bf M}+{\bf J}).\tag66b$$

The projected constraint
functions are distinguished by the fact that their Hamiltonian vector fields
are complete; in particular, the finite transformations generated by these
functions map spacelike hypersurfaces to spacelike hypersurfaces.  For this
reason,
despite the technical complexity of the algebraic structure involved, one
may choose to view the constraints \(64) and the ``hypersurface deformation
algebra'' \(66b) as
fundamental.  At any rate, there is no known alternative to this ``open
algebra'' in general relativity.  Consequently, it is of interest to interpret
this
algebra from the perspective of the covariant phase space.

Because there
are no hypersurfaces to be found in the realm of the covariant phase space,
in order to make contact with the algebra \(66b) some spacelike
hypersurfaces
will have
to be provided.  So, for the purposes of the present discussion, let us
assume that we identify $\Mm=R\times\Sigma$ and require that the
diffeomorphisms
$X$ are in fact spacelike foliations\footnote*{If desired, a fixed spacelike
foliation $Y:R\times\Sigma\to\Mm$ can be introduced, and points in the
covariant phase space identified with pairs $(Y^*\phia, X\circ Y)$.}
$X:R\times\Sigma\to\Ma$.  Each
foliation $X$
provides $\Ma$ with an adapted hypersurface basis built from
the unit
normal and tangent vectors to the hypersurfaces in $\Ma$, which are
defined as in
\S2.  In  particular
$$\eqalign{n_\alpha X^\alpha_a&=0\cr
g^{\alpha\beta}n_\alpha n_\beta &= -1.}\tag67$$
A basis on $\Mm$ can then be obtained by pull-back from that on
$\Ma$, the
relationship being
$$\eqalign{n^\mu&=X^\mu_\alpha n^\alpha\circ X\cr
X^\mu_a&=X^\mu_\alpha X^\alpha_a\circ X .}\tag68$$

On $\Gamma$, the hypersurface deformations can be viewed as a
modification of the
{\it right} action of the infinitesimal diffeomorphisms that is available
when
one has a spacelike foliation.  We build a vector field $N^\mu$ on
$\Mm=R\times\Sigma$ by specifying the amount of normal ($N^\sperp$)
and tangential ($N^a$)
deformation of each hypersurface $\Sigma$:
$$N^\mu = N^\sperp n^\mu + N^a X^\mu_a.\tag69$$
To compute the induced action of these vector fields on the covariant phase
space
we consider the pure gauge
vector fields ${\cal Z, Z^\prim}$ on $\Gamma$
defined by
${\cal Z}=(L_N\phia, L_N X), {\cal Z^\prim}=(L_{M}\phia,
L_{M}
X)$, ($M^\mu$ is
defined similarly to $N^\mu$) and compute the commutator
$[{\cal Z, Z^\prim}]$.  The variations in $n^\mu$ are computed using
\(67), \(68); after a straightforward computation we find
$$[{\cal Z, Z^\prim}]={\cal Z^{\prim\prim}},\tag610$$
where
$${\cal Z^{\prim\prim}}=(L_{N\cdot M}\phia, L_{N\cdot
M}X),$$
$$\eqalign{(N\cdot M)^\mu=\Big(&M^{ a}X^\nu_a\nabla_\nu
N^\sperp-
N^aX^\nu_a\nabla_\nu M^{\sperp}
\Big)n^\mu\cr
&+ \Big(\gamma^{ab}X^\nu_b(M^\sperp\nabla_\nu N^{\sperp}-
N^{\sperp}\nabla_\nu M^\sperp) + (M^b\nabla_b N^a-
N^b\nabla_b M^a)\Big)X^\mu_a,}\tag611$$
and $\gamma^{ab}$ is the inverse metric induced on the leaves of the
foliation.
Comparing \(611) with \(66b) it follows that the construction $(N^\sperp,
N^a)\to
N^\mu\to{\cal Z}$
described above represents an anti-homomorphism from the algebra of
hypersurface deformations into the algebra of ($\Omega$ preserving)
vector fields on $\Gamma$.
\subhead{\it Observables}
In the canonical Hamiltonian formulation of dynamical systems with (first
class)
constraints ``observables'' are defined as functions on the phase space that
have a vanishing Poisson bracket with the constraint functions modulo the
constraints.  More geometrically, observables are functions on the phase
space which project to the space of orbits of the Hamiltonian vector fields
in the constraint surface.  These abstract ways of defining observables are
meant to capture the notion of observables as ``gauge invariant'' functions
on the physically accessible portion of the phase space.

In the canonical formulation of parametrized field theories the constraint
functions generate canonical transformations corresponding to the change
in the phase space data as the hypersurface they are on is deformed through
spacetime.  The embeddings change according to the deformation, the truly
dynamical variables $(q^A, \Pi_A)$ change according to the dynamical
equations, and this induces the change in the embedding momenta via the
constraints \(215) which are preserved in the course of the dynamical
evolution.
Because this motion on $\Upsilon$ can be viewed as the infinitesimal action
of \DiffMtxt \refto{Isham1985}, it follows that in the canonical formalism the
observables can
be
equivalently characterized as either constants of motion, or invariants
under infinitesimal diffeomorphisms.  The latter characterization makes
direct contact with the covariant phase space notion of observables, but the
observables constructed in \S5 differ somewhat from the constant of
motion observables on $\Upsilon$.

To see this we must spell out the construction of observables in
canonical parametrized field theory, which is essentially an application of
Hamilton-Jacobi theory.  Imagine solving the Hamilton equations of motion
for the canonical variables.  This can be done by solving the
many-fingered time functional differential equations \refto{Kuchar1972}
$$\eqalign{{\delta q^A(x)\over\delta
Q^\alpha(y)}&=[q^A(x),H_\alpha(y)]\cr
 {\delta \Pi_A(x)\over\delta
Q^\alpha(y)}&=[\Pi_A(x),H_\alpha(y)]}\tag612$$
along with the constraints \(215).  Here the brackets are the Poisson
brackets, and we allow the usual abuse of notation which identifies the
solutions to the equations of motion with the canonical variables
themselves.  The solution is thus specified by giving the
canonical data as a functional of the embeddings $Q$ and a set of initial
data $(\hat q^A, \hat\Pi_A)$ on an initial embedding $\hat Q$:
$$\eqalign{q^A &= q^A[Q,\hat q^A,\hat\Pi_A]\cr
\Pi_A &= \Pi_A[Q, \hat q^A,\hat\Pi_A],}\tag613$$
$$\eqalign{q^A[\hat Q,\hat q^A,\hat\Pi_A]&=\hat q^A\cr
\Pi_A[\hat Q, \hat q^A,\hat\Pi_A]&=\hat \Pi_A.}\tag614$$
For each $Q$ (and $\hat Q$) eqs. \(613) specify a diffeomorphism $(\hat
q^A,
\hat\Pi_A)\longrightarrow(q^A,\Pi_A)$ that preserves the natural
symplectic structure on the space of pairs $(q^A,\Pi_A)$.  In other words,
dynamical evolution is a canonical
transformation.  Inverting the map \(613) amounts to expressing the initial
data as a functional of the solution:
$$\eqalign{\hat q^A &= \hat q^A[Q,  q^A,\Pi_A]\cr
\hat\Pi_A &= \hat\Pi_A[Q, q^A,\Pi_A],}\tag615$$
Because initial data are always ``constants of the motion'', the functionals
on $\Upsilon$ specified in \(615) will have (strongly) vanishing Poisson
brackets with the constraint functions,
$$[\hat q^A,H_\alpha]=0=[\hat\Pi_A,H_\alpha],\tag(616)$$
and therefore represent a set of observables.  Obviously, this set is
complete.  Therefore in canonical parametrized theory the natural
observables correspond to the freely specifiable Cauchy data
$(\hat q^A,\hat\Pi_A)$ on a hypersurface determined by $\hat Q$.  Note
that this
means that there will always be a symplectic diffeomorphism which
identifies the observables with points in the canonical phase space for the
unparametrized theory.

Our construction of the observables ${\cal O}^A$ on the covariant phase
space also led back to the unparametrized theory: the space of observables
is equivalent to the space of solutions to \(22).  As mentioned above, the
space
of solutions to \(22) is symplectically diffeomorphic to the space of Cauchy
data for \(22) which is, in turn, (assumed) equivalent to the canonical
phase space of the unparametrized theory.  Thus the reduced phase spaces
in each case coincide: $\hat\Gamma\simeq\hat\Upsilon$.

Notice however that it is {\it only} the reduced phase spaces which
coincide.  The relation between $\Gamma$ and $\Upsilon$ is not entirely
simple; indeed, one can at best identify $\bar\Upsilon$ with an open subset
of $\Gamma$.  In particular, $\Gamma$ admits an action of \DiffMtxt\
but $\Upsilon$ (or $\bar\Upsilon$) does not.  Thus, while $\hat \Gamma$
arises as $\hat\Gamma=\Gamma/\DiffM$, $\hat\Upsilon$ is obtained as the
space of orbits of a much more complicated structure than a Lie group and
these orbits are in a rather different space than $\Gamma$.  In
this sense it is perhaps remarkable that $\hat\Gamma\simeq\hat\Upsilon$.
\vskip 0.5truein
I would like to thank Karel Kucha\v r for discussions.
\references
\refis{ADM1959}{R. Arnowitt, S. Deser, and C. Misner, \pr 116, 1322,
1959.}

\refis{ADM1962}{R. Arnowitt, S. Deser, and C. Misner in {\it
Gravitation: An
Introduction to Current Research}, edited by L. Witten (Wiley, New York
1962).}

\refis{Arms1979}{J. Arms, \jmp 20, 443, 1979.}

\refis{Arms1980}{J. Arms, \jmp 21, 15, 1980.}

\refis{Ashtekar1991a}{A. Ashtekar, {\it Lectures on Non-Perturbative
Canonical
Gravity}, (World Scientific, Singapore 1991) and references therein.}

\refis{Ashtekar1991b}{ A. Ashtekar, L. Bombelli, and O.
Reula, in {\it Mechanics,
Analysis and Geometry : 200 Years After Lagrange}, edited by M.
Francaviglia ( North-Holland, New York 1991). }

\refis{Dirac1964}{P.A.M. Dirac, {\it Lectures on Quantum Mechanics},
(Yeshiva
University, New York, 1964).}

\refis{Fischer1979}{See A. Fischer and J. Marsden in {\it General
Relativity: An
Einstein Centenary Survey}, edited by S. Hawking and W. Israel
(Cambridge University Press, Cambridge 1979).}

\refis{Halliwell1991}{J. Halliwell, \prd 43, 2590, 1991.}

\refis{Henneaux1991}{G. Barnich, M. Henneaux, C. Schomblond, \prd 44, R939,
1991.}

\refis{Hormander1966}{L. H\" ormander, \journal Ann. Math., 83, 129,
1966.}

\refis{Isenberg1982}{J. Isenberg and J.
Marsden, \prpts 89, 181, 1982, and references therein.}

\refis{Isham1985}{C. J. Isham and K. V. Kucha\v r, \ann 164, 288, 1985;
\ann 164, 316, 1985.}

\refis{Kuchar1971}{K. V. Kucha\v r, \prd 4, 955,
1971.}

\refis{Kuchar1972}{K. V. Kucha\v r, \jmp 13, 758, 1972}.

\refis{Kuchar1976}{K. V. Kucha\v r, \jmp 17, 801, 1976.}

\refis{Kuchar1978}{K. V. Kucha\v r, \jmp 19, 390, 1978.}

\refis{Kuchar1992}{K. V. Kucha\v r, ``Time and Interpretations of
Quantum Gravity'', to
appear in the proceedings of {\it The Fourth Canadian Conference on
General Relativity and Relativistic Astrophysics}, edited by G. Kunstatter,
D. Vincent, and J. Williams (World Scientific, Singapore 1992).}

\refis{Lanczos1970}{C. Lanczos, {\it The Variational Principles of
Mechanics} (University of Toronto Press, Toronto 1970).}

\refis{CGT1989}{K. V. Kucha\v r, C. G. Torre, \jmp 30, 1769, 1989.}

\refis{CGT1990}{K. V. Kucha\v r and C. G. Torre, \prd 43, 419, 1990.}

\refis{CGT1991a}{K. V. Kucha\v r and C. G. Torre, \prd 44,
3116, 1991.}

\refis{CGT1991b}{K. V. Kucha\v r and C. G. Torre in {\it Conceptual
Problems
of Quantum Gravity}, edited by A. Ashtekar and J. Stachel, (Birkh\"auser,
Boston 1991).}

\refis{CGT1991c}{ C. G. Torre, \cqg 8, 1895, 1991.}

\refis{CGT1992a}{C. G. Torre, ``Is General Relativity an `Already
Parametrized' Theory?'', Utah State University Preprint, 1992.}

\refis{CGT1992b}{C. G. Torre, ``Covariant Phase Space Formulation of
Parametrized Field Theories'', Utah State University Preprint, 1992.}

\refis{CGT1992c}{C. G. Torre, in preparation, 1992.}

\refis{Wald1990}{J. Lee and R. Wald, \jmp 31, 725, 1990.}

\refis{Witten1987}{C. Crnkovic and E. Witten in {\it 300 Years of
Gravitation},
edited by S. Hawking and W. Israel (Cambridge University Press,
Cambridge 1987).}

\refis{York1980}{Y. Choquet-Bruhat and J. York in {\it General
Relativity
and Gravitation: 100 Years After the Birth of Albert Einstein, Vol. 1},
edited by A. Held (Plenum, NY 1980).}

\endreferences
\endit